\documentclass[preprint,showpacs,preprintnumbers,amsmath,amssymb]{revtex4}
\usepackage{graphicx}
\usepackage{dcolumn}
\usepackage{bm}

\begin{document}

\title{Spin-dependence of Ce $4f$ hybridization in magnetically ordered systems: A spin-resolved photoemission study of Ce/Fe(110)}

\author{Yu. S. Dedkov,$^{1,}$\footnote{Corresponding author. E-mail: dedkov@physik.phy.tu-dresden.de} M. Fonin,$^2$ Yu. Kucherenko,$^3$ S. L. Molodtsov,$^1$ U. R\"udiger,$^2$ and C. Laubschat$^1$}
\affiliation{ 
 $^1$Institut f\"ur Festk\"orperphysik, Technische Universit\"at Dresden, 01062 Dresden, Germany\\
 $^2$Fachbereich Physik, Universit\"at Konstanz, 78457 Konstanz, Germany\\
 $^3$Institute for Metal Physics, National Academy of Sciences of Ukraine, 03142 Kiev, Ukraine}

\date{\today}

\begin{abstract}
Spin- and angle-resolved resonant (Ce $4d\rightarrow4f$) photoemission spectra of a monolayer Ce on Fe(110) reveal spin-dependent changes of the Fermi-level peak intensities. That indicate a spin-dependence of $4f$ hybridization and, thus, of $4f$ occupancy and local moment. The phenomenon is described in the framework of the periodic Anderson model by $4f$ electron hopping into the exchange split Fe $3d$ derived bands that form a spin-gap at the Fermi energy around the $\overline{\Gamma}$ point of the surface Brillouin zone.       
\end{abstract}


\maketitle

As a function of chemical composition, the electronic properties of Ce $4f$ states in intermetallic compounds vary from localized $4f^1$ character over heavy-fermion behavior and mixed valence to the boarder of itinerant behavior~\cite{Sereni:1991}. This fascinating variety of characters is already reflected in Ce metal, where in the course of the famous isostructural $\gamma\rightarrow\alpha$ transition a magnetic phase transforms into a nonmagnetic one depending on temperature and/or pressure accompanied by a volume collapse of 15\%~\cite{Koskenmaki:1978}. While in the promotion model this phenomenon was ascribed to a transition from a trivalent $4f^1(5d6s)^3$ to a tetravalent $4f^0(5d6s)^4$ configuration~\cite{Ramirez:1971}, later studies related the effect to a Mott-transition from localized to itinerant character of the $4f$ state~\cite{Johansson:1974} or to a Kondo collapse~\cite{Allen:1982}.

The promotion model is clearly ruled out by photoemission (PE) that reveals only weak intensity changes of the total $4f$ derived emission upon the $\gamma\rightarrow\alpha$ transition~\cite{Wieliczka:1982}. Instead of a single $4f^0$ PE final state at about 2\,eV binding energy (BE) as expected from a localized $4f^1$ ground state a second $4f$-derived feature is observed at the Fermi energy, $E_F$, that increases in intensity upon the $\gamma\rightarrow\alpha$ transition~\cite{Wieliczka:1982}. An itinerant description based on the local density approximation (LDA) fails to explain this double-peak structure~\cite{Note:DMFT}, it is, however, well reproduced in the framework of the single-impurity Anderson model (SIAM) considering electron hopping between localized $4f^1$ and valence-band (VB) states~\cite{Gunnarsson:1983}. A momentum dependence of the 4f signal as recently observed by angle-resolved PE experiments~\cite{Andrews:1996,Danzenbacher:2005,Vyalikh:2006} could be explained considering the translational symmetry of the solid within a simple approach to the periodic Anderson model (PAM)~\cite{Danzenbacher:2005,Vyalikh:2006,Danzenbacher:2006}.

From both SIAM and PAM the Fermi-peak intensity may be taken as a direct measure for the hopping probability. The latter should increase with the VB density of states at $E_F$, and in fact huge Fermi-level peaks are typically observed in PE spectra of Ce transition-metal compounds reflecting $\alpha$-like behavior of the Ce $4f$ states due to hybridization with transition-metal $d$-bands~\cite{Jung:2003}. A spin-dependence of $4f$ hopping may be expected for magnetically ordered systems where the exchange splitting of the VB leads to strong variations of the density of states at $E_F$ for differently oriented VB spins. Respective spin-dependent $\gamma\rightarrow\alpha$ transitions have not be observed so far, the effect, however, could be of high importance for the understanding of magnetic anomalies in these systems since the local magnetic properties of the Ce atoms may strongly vary as a function of $4f$ spin orientation.

In this contribution we report for the first time on a spin-dependent $\gamma\rightarrow\alpha-$like transition observed by a spin- and angle-resolved resonant PE from an ordered Ce adlayer on Fe(110). Although hybridization is expected to be relatively weak in the outermost surface layer due to the low coordination of the Ce atoms~\cite{Laubschat:1990}, the quasi two-dimensional structure of the system allows for a proper determination of the position in $\mathbf{k}$ space probed in the experiment as necessary for a quantitative description within PAM applied here. For Ce/Fe(110), our local spin density approximation (LSDA) slab calculations reveal at the $\overline{\Gamma}$ point a strong reduction of majority-spin states around $E_F$ that should lead to a respective weakening of $4f$ hybridization for this spin orientation. In fact, our spin- and angle-resolved PE spectra show a lower Fermi-level peak intensity for the $4f$ majority- than minority-spin orientation. Simulations of the PE spectra within PAM reproduce this effect as well as a spin-dependent splitting of the ionization peak observed in the experimental data. Similar spin-dependencies are expected to be of high importance for the understanding of magnetic anomalies in a series of other RE systems, where hybridization phenomena were experimentally observed and successfully described  within SIAM or PAM~\cite{Danzenbacher:2006,Kucherenko:2002}.  

\begin{figure}[tb]
\vspace{0pt}
\hspace{-20pt}
\includegraphics[scale=1.0]{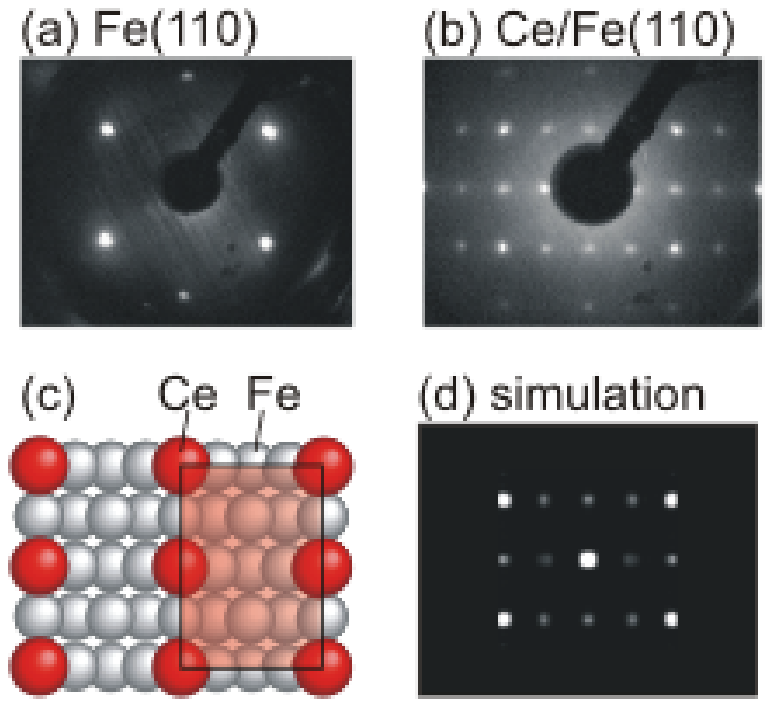}
\vspace{0pt}
\caption{\label{fig1} (Color online) LEED images obtained from (a) Fe(110) and (b) Ce/Fe(110); assumed surface crystallographic structure of the Ce/Fe(110) system (c) and simulation of the LEED-image (d). The shaded rectangle in (c) visualizes the \textit{fcc} Ce(110) plane expanded by 11\%.}
\end{figure}

A Fe(110) substrate was prepared by thermal deposition of Fe films with a thickness of 50\,\AA\ on W(110) and subsequent annealing at 450\,K. Low-energy electron diffraction (LEED) yielded in sharp patterns with two-fold symmetry as expected for a structurally ordered \textit{bcc} Fe(110) surface [Fig.~\ref{fig1}(a)]. Further deposition of 0.5 monolayer (close-packed atomic arrangement) of Ce metal at 300\,K led to a sharp overstructure in the LEED pattern [Fig.~\ref{fig1}(b)] that could be reproduced by a kinematic LEED simulation [Fig.~\ref{fig1}(d)] with the structural model shown in Fig.~\ref{fig1}(c). Ce atoms are placed on hollow-sites of the \textit{bcc} Fe(110) surface reproducing the arrangement of a (110) plane of \textit{fcc} $\gamma$-Ce expanded by 11\%. Spin- and angle-resolved resonant PE experiments at the Ce $4d\rightarrow4f$ absorption threshold were performed using a hemisherical PHOIBOS\,150 electron-energy analyzer (SPECS) equipped with a 25\,kV mini-Mott spin-detector and synchrotron radiation from beamline U125/1-PGM of BESSY (Berlin). The energy and angle resolutions were set to 100\,meV and $\pm2^\circ$, respectively. The light incidence angle was 30$^\circ$ with respect to the sample surface, and the photoelectrons were collected around the surface normal. Spin-resolved measurements were performed in normal emission geometry at 130\,K in magnetic remanence after having applied a magnetic field pulse of about 500\,Oe along the in-plane $\left\langle 1\bar{1}0 \right\rangle$ easy axis (perpendicular to electric field vector of the light) of the Fe(110) film. The experimental setup asymmetry was accounted for in the standard way by measuring spin-resolved spectra for two opposite directions of applied magnetic field~\cite{Kessler:1985,Johnson:1992}. The base pressure in the experimental chamber was in the upper 10$^{-11}$\,mbar range rising shortly to the upper 10$^{-10}$ range during evaporation and annealing.

\begin{figure}[t]
\vspace{-55pt}
\hspace{-30pt}
\includegraphics[scale=0.5]{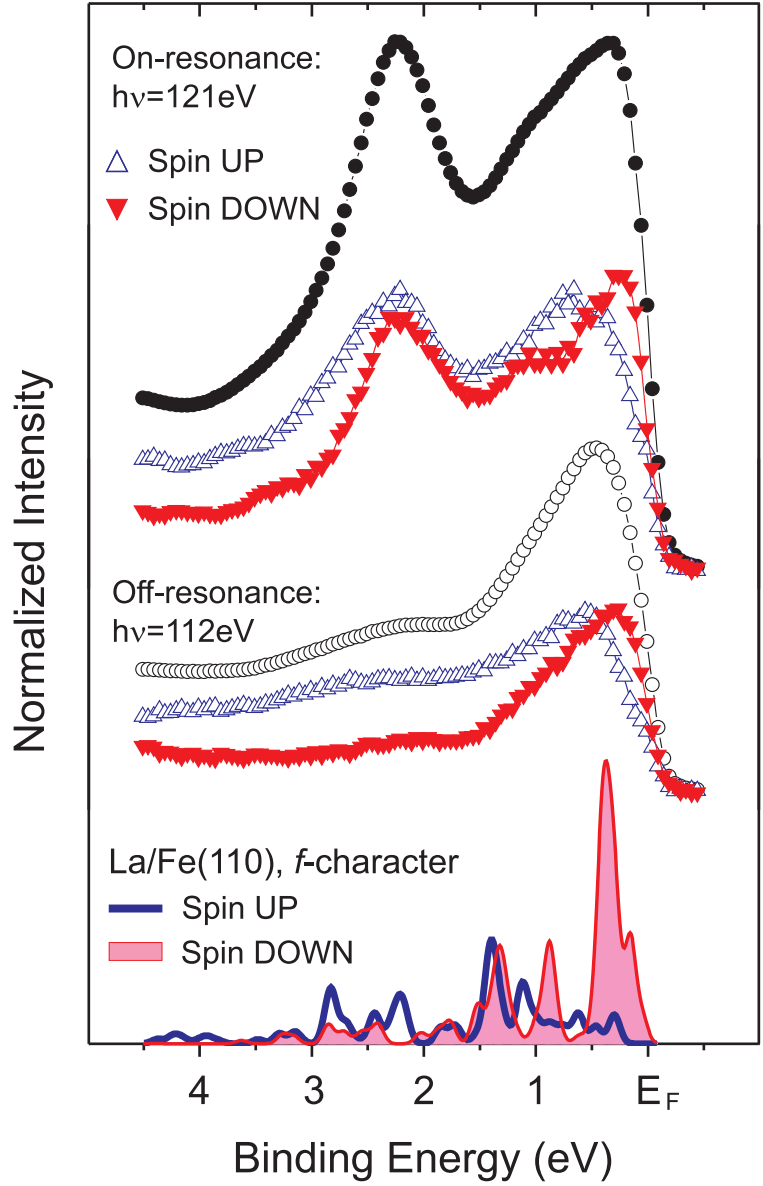}
\vspace{-50pt}
\caption{\label{fig2} (Color online) Spin-resolved PE spectra of Ce/Fe(110) system measured in on- and off-resonance at the $4d\rightarrow4f$ absorption threshold. Open/filled triangles denote contributions of majority/minority spin directions, respectively. Bottom part: Calculated local $4f$ character of the VB states ($|c_f^{\sigma}(E,\Gamma)|^2$) at the La site in the $\Gamma$ point of the surface BZ for La/Fe(110) system for majority- (solid line) and minority-spin (shaded area) direction.}
\end{figure}

Fig.~\ref{fig2} shows spin-resolved PE data of Ce/Fe(110) taken on- and off-resonance at 121\,eV and 112\,eV photon energies, respectively. The off-resonance spectra are dominated by emissions from Fe $3d$-derived bands and are very similar to respective data of the pure Fe substrate (not shown here). The spectra reflect clearly the exchange splitting of the Fe $3d$ bands into a minority-spin component at E$_F$ ("spin down": filled triangles) and a majority-spin component shifted to higher BE ("spin up": open triangles). While the spectra of the pure substrate remain almost unchanged when going from 112\,eV to 121\,eV photon energy, the on-resonance spectra of Ce/Fe(110) reveal an additional feature around 2.2\,eV BE that is ascribed to the resonantly enhanced $4f$ signal.
 
\begin{figure}[b]
\vspace{-30pt}
\includegraphics[scale=0.7]{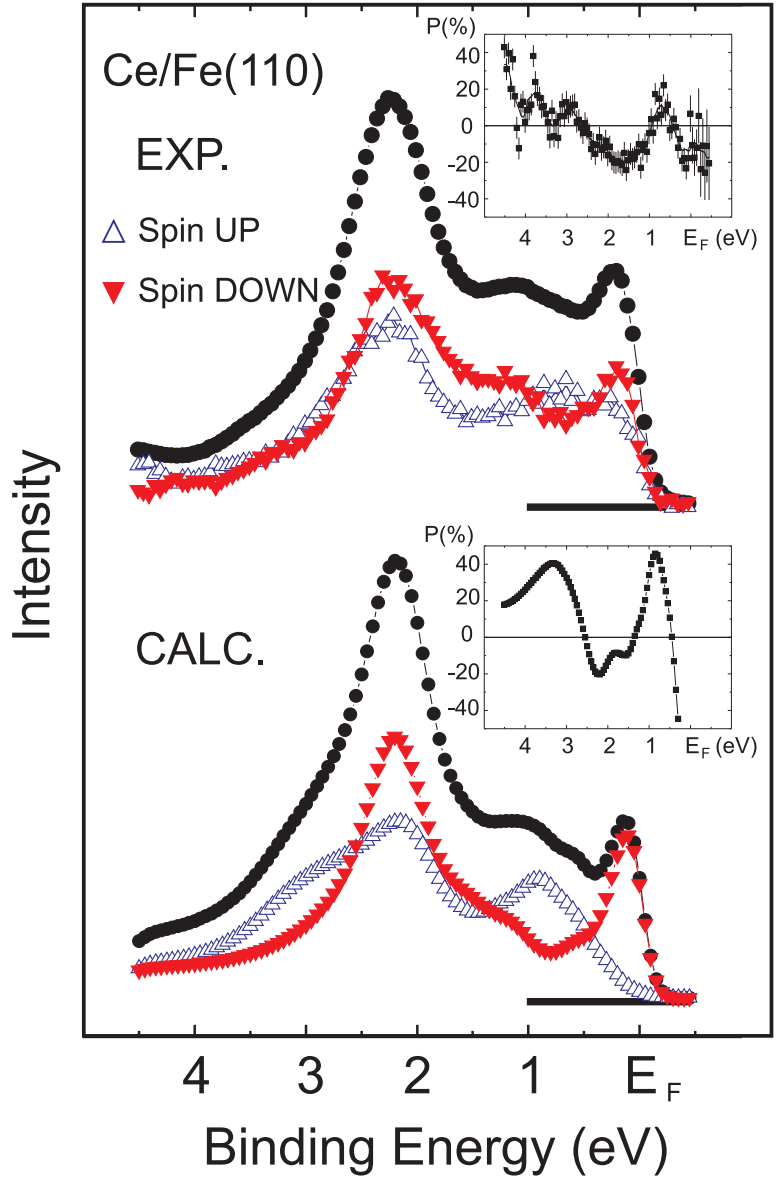}
\vspace{-60pt}
\caption{\label{fig3} (Color online) Spin-resolved experimental (upper part) and calculated (lower part) Ce $4f$ emission for Ce/Fe(110). Majority- and minority-spin emissions are shown by open and solid triangles, respectively. The insets show the corresponding spin polarization $P$.}
\end{figure}

In order to extract the Ce $4f$ contributions from these spectra, the off-resonance data were subtracted from the on-resonance spectra after proper normalization of the intensities with respect to the photon flux and the slowly varying Fe $3d$ photoionization cross section. The resulting spin-resolved $4f$ spectra are shown in the upper part of Fig.~\ref{fig3} together with the corresponding spin polarization $P$ (inset) defined as $P=(I^\uparrow-I^\downarrow)/(I^\uparrow+I^\downarrow)$, where $I^\uparrow$ and $I^\downarrow$ denote the intensities of the majority- and minority-spin channels, respectively. The spectra reveal the well-known double-peak structure of the Ce $4f$ emission consisting of a main maximum at 2.2\,eV corresponding to the ionization peak expected for an unhybridized $4f^1$ ground state and the hybridization peak at E$_F$. From the weak intensity of the latter relative to the ionization-peak signal, a weak hybridization similar to the one in $\gamma$-Ce can be concluded as it is expected for a Ce surface layer~\cite{Weschke:1991}. The most important observation is, however, that the intensity of the hybridization peak is larger for the minority- than for the majority-spin component (Fig.~\ref{fig3}) indicating larger $4f$-hybridization of the former. The spin polarization of both, the ionization and the hybridization peaks, gives a negative sign indicating that the preferred orientation of the Ce $4f$ spins is opposite to the magnetization direction of the Fe layers. In addition to the double-peak structure another feature is visible around 1\,eV BE (Fig.~\ref{fig3}), that is weaker in intensity and shifts to lower BE when going from the minority- to the majority-spin component.
 
In order to understand the ground-state magnetic properties of Ce/Fe(110), as a first step fully relativistic spin-polarized band-structure calculations were performed by means of the linear muffin-tin orbital (LMTO) method. A pure Fe surface and the Ce/Fe(110) system were considered using the structural model shown in Fig.~\ref{fig1}(c). The Fe substrate was simulated by a five-layer slab of Fe atoms with (110) orientation of the surface. The results were compared to data calculated for the isostructural non-$f$ system La/Fe(110).

For the atoms in the middle layer of the Fe slab the calculations give a local electronic structure close to that obtained for Fe bulk~\cite{Chasse:2003}. The calculated Fe $3d$ spin moment value lies between 2.35\,$\mu_B$ and 2.40\,$\mu_B$. At the surface it increases to 2.60\,$\mu_B$. In all cases contributions of $s$ and $p$ electrons to magnetic moment are negligible.

By the presence of a Ce overlayer the Fe $3d$ spin moments of the surface atoms are reduced to 2.14\,$\mu_B$ and 2.50\,$\mu_B$, respectively, depending on whether the Fe atoms are nearest neighbors of Ce atoms or not. Replacing in the calculation Ce by La atoms give very similar results indicating that the electronic structure of the Fe atoms is perturbed by interactions with extended valence states (mainly $5d$) of the overlayer.

The calculations yield for a La atom on the Fe(110) surface a local spin moment of $-0.24\,\mu_B$, determined mainly by the $5d$ electrons ($-0.20\,\mu_B$). The negative sign stands for an antiparallel orientation with respect to the Fe $3d$ spin moment. For the Ce atom the local spin moment is equal to $-1.12\,\mu_B$, with $5d$ and $4f$ contributions of $-0.28\,\mu_B$ and $-0.82\,\mu_B$, respectively. Thus, like in other Ce-Fe systems~\cite{Finazzi:1995,Arend:1998,Konishi:2000}, the Ce $4f$ electrons reveal a spin orientation opposite to Fe $3d$ majority spin in agreement with the PE experiment. Since the $4f$ electrons have additionally a large positive orbital momentum of 2.80\,$\mu_B$ due to their reduced atomic coordination at the surface the total moment equals to 1.70\,$\mu_B$ and corresponds, thus, to ferromagnetic coupling with respect to the Fe $3d$ spins. At finite temperatures magnetic disorder leads to the situation encountered in the experiment where a part of the $4f$ spins are flipped into the opposite direction.

In order to describe the observed variation of $4f$ hybridization as a function of spin orientation, we used the simplified periodic Anderson model that was recently successfully applied to explain the angle-resolved PE spectra of CePd$_3$~\cite{Danzenbacher:2005} and Ce/W(110)~\cite{Vyalikh:2006}. In this approach the double occupation of the $4f$ states is ignored (on-site $f-f$ Coulomb interaction energy, $U_{ff}\rightarrow\infty$) and ${\mathbf k}$ vector conservation upon hybridization is assumed. In this case a simplified (without $U_{ff}$ term) Anderson Hamiltonian can be written as follows
\begin{eqnarray}
H &=& \sum_{{\bf k},\sigma} \varepsilon^{\sigma}({\bf k})
           d_{{\bf k}\sigma}^{+} d_{{\bf k}\sigma}^{}
     + \sum_{{\bf k},\sigma} \varepsilon_f^{\sigma}({\bf k})
           f_{{\bf k}\sigma}^{+} f_{{\bf k}\sigma}^{}  \nonumber\\
  && + \sum_{{\bf k},\sigma} V_{\bf k}^{\sigma}(E)
          \left(d_{{\bf k}\sigma}^{+}f_{{\bf k} \sigma}^{}+
               f_{{\bf k} \sigma}^{+}d_{{\bf k}\sigma}^{}\right), \nonumber
\end{eqnarray}
where the VB states $|{\bf k}\sigma\rangle$ have a dispersion $\varepsilon^{\sigma}({\bf k})$ and are described by creation (annihilation) operators $d_{{\bf k}\sigma}^{+}$ ($d_{{\bf k}\sigma}^{}$). The operator $f_{{\bf k}\sigma}^{+}$ creates a $f$ electron with momentum ${\bf k}$, spin $\sigma$, and energy $\varepsilon_f^{\sigma} ({\bf k})$. We assume that a non-hybridized $f$ band has no dispersion:
$\varepsilon_f^{\sigma} ({\bf k})=\varepsilon_f^{\sigma}$ allowing, however, a possible small difference in the energy positions of $4f$ levels with different spin $\sigma$ due to exchange interaction. The two electron subsystems (VB and $4f$ states) are coupled via a hybridization $V_{\bf k}^{\sigma}(E)$ that leaves the electron spin unaffected, i.\,e. spin-flips upon electron hopping are excluded. $E$ denotes the BE with respect to E$_F$. This form of the Hamiltonian allows us to diagonalize it for each particular ${\bf k}$ point of the surface Brillouin zone (BZ) and for each spin state $\sigma$.

For the hybridization matrix element $V_{\bf k}^{\sigma}(E)$ we use calculated $f$-projected local expansion coefficients $c_f^{\sigma}(E,{\bf k})$ of the Bloch functions
around the rare-earth sites: $V_{\bf k}^{\sigma}(E)=\Delta\cdot c_f^{\sigma}(E,{\bf k})$, where $\Delta$ is a constant, adjustable parameter. Expansion coefficients $c_f^{\sigma}(E,{\bf k})$ that characterize the local $f$ character of VB states were taken from the results of the band-structure calculations of the La/Fe(110) system, in order to exclude the contribution of localized Ce $4f$ orbitals. For normal emission of the photoelectrons we have to consider VB states at the $\Gamma$ point of the surface BZ. The calculated values of $\left|c_f^{\sigma}(E,\Gamma)\right|^2$ are shown in the bottom part of Fig.~\ref{fig2}. The energy distributions of the VB states of local $f$ character are quite different for majority- and minority-spin electrons. Since these states are formed by linear combination of wave functions of the neighboring atoms (mainly Fe $3d$) penetrating into the La atomic spheres, they reflect to some extent the energy and spin distribution of the latter (see off-resonance spectra in Fig.~\ref{fig2}). Their different amplitude and energy distribution for majority- and minority-spin states causes strong differences in the respective hybridization matrix elements and results in different shape of the $4f$ PE spectra for the two spin directions.

The spectral functions of the Ce $4f$ emission were calculated using the parameters $\varepsilon_f^{\uparrow}=-1.9$\,eV, $\varepsilon_f^{\downarrow}=-1.7$\,eV, and $\Delta=0.85$\,eV. These values deviate from those used in Ref.~\cite{Vyalikh:2006} for Ce/W(110) only by slightly higher BE of the non-hybridized $4f$ level resulting from the lower coordination of the Ce atoms. An energy-dependent life-time broadening of the form $\Gamma_L =0.030$\,eV$+0.085E$ was considered. The calculated spectral functions were additionally broadened with a Gaussian ($\Gamma_G=100$\,meV) to simulate finite instrumental resolution and an integral background was added to take into account inelastic scattering.

The calculated spin-resolved Ce $4f$ PE spectra are presented in Fig.~\ref{fig3} (lower part). The energy distribution of the PE intensity agrees well with that of the experimental spectra (Fig.~\ref{fig3}, upper part). The minority-spin spectrum reveals high intensity of the hybridization peak due to large density of the minority-spin VB states close to E$_F$. A shoulder near 1\,eV BE is formed by hybridization with VB peaks at 0.9\,eV and 1.3\,eV BE (Fig.~\ref{fig1}). In accordance with the experiment, in the calculated majority-spin spectrum the ionization peak is split into three components (maxima at 0.9\,eV, 2.1\,eV, and shoulder at 3\,eV BE) as a result of hybridization with the VB states (at 1.4\,eV and between 2\,eV and 3\,eV BE). No majority-spin hybridization peak is obtained in the calculation due to the negligibly small density of VB states for this spin direction at the Fermi level. This theoretical result deviate from the experiment where a reduced but finite hybridization peak was observed. The latter may be ascribed to the finite angle resolution of the experiment that samples also regions in the $\mathbf{k}$ space where majority-spin bands cross E$_F$. The calculated spin polarization (Fig.~\ref{fig3}, inset in the lower part) reproduces qualitatively the energy dependence of the measured polarization. Particularly good agreement is obtained for the points where the spin polarization changes its sign. 

\textit{In summary}, we have shown that the observed spin-dependence of the shape of the Ce $4f$ emission in Ce/Fe(110) system may be explained by a spin-dependence of $4f$-hybridization. From this result $4f$-occupancy as well as effective magnetic moment are generally expected to vary with spin-orientation, an effect that may be of crucial importance for the understanding of many-body effects and magnetic anomalies in RE systems.

This work was funded by the Deutsche Forschungsgemeinschaft, SFB 463, Projects TP B4 and TP B16 as well as SFB513. We would like to acknowledge BESSY staff for technical support during experiment.

\clearpage

\clearpage

\clearpage

\end{document}